\documentclass{svproc}
\usepackage{graphicx}%
\usepackage{multirow}%
\usepackage{amsmath,amssymb,amsfonts}%
\usepackage{mathrsfs}%
\usepackage[title]{appendix}%
\usepackage{xcolor}%
\usepackage{textcomp}%
\usepackage{manyfoot}%
\usepackage{booktabs}%
\usepackage{algorithm}%
\usepackage{algorithmicx}%
\usepackage{algpseudocode}%
\usepackage{listings}%
\usepackage{url}

\begin{document}
\mainmatter              % start of a contribution
\title{Phase Boundary of a Stochastic Watts-Threshold SIS Model on Random Networks}
\titlerunning{Phase Boundary of a Watts-Threshold SIS Model}  % abbreviated title (for running head)
\author{Yasmine Beji\inst{1,3} \and Heger Arfaoui\inst{2,3} \and 
Slimane BenMiled \inst{3} }
\authorrunning{Yasmine Beji et al.} % abbreviated author list (for running head)

\institute{Mediterranean Institute of Technology, South Mediterranean University\\
\and
\'Ecole Nationale d'Ing\'enieurs de Tunis, Universit\'e Tunis El Manar \\
\and
BIMS Lab, Institut Pasteur de Tunis}

\maketitle              % typeset the title of the contribution

\begin{abstract}
Complex contagion models, in which adoption requires reinforcement from multiple neighbors, have been extensively studied in the monotone (no-recovery) setting, but the phase diagram of threshold models with SIS-like recovery on networks remains unmapped. We study a stochastic Watts-threshold SIS model on Erd\H{o}s--R\'enyi and Barab\'asi--Albert networks and reconstruct its extinction--persistence phase boundary in the joint parameter space of transmission rate~$\beta$, adoption threshold~$\theta$, and infectious duration~$d$. Using adaptive Delaunay-based sampling and weighted logistic regression on over 180,000 Monte Carlo trials, we find that: (i)~the boundary is well described by a six-parameter interaction model whose structure is invariant across both topologies; (ii)~the transition is sharp, with the 10--90\% extinction-probability band spanning only $\Delta\theta \approx 0.005$--$0.008$; and (iii)~the adoption threshold is the dominant parameter governing epidemic feasibility, with transmission rate and infectious duration playing secondary and asymmetric roles. The characterization provides a quantitative reference for the complex-contagion analogue of the classical SIS epidemic threshold.
\end{abstract}
%
% ================================================================
\section{Introduction}

In classical SIS epidemic models on networks, the persistence of an epidemic is governed by a single compound quantity: the basic reproduction number $R_0 = \beta \, d \, \langle k \rangle$, where $\beta$ is the per-contact transmission probability, $d$ the mean infectious duration, and $\langle k \rangle$ the mean degree~\cite{pastor-satorras2015}. When $R_0 > 1$ the epidemic persists at an endemic equilibrium; when $R_0 < 1$ it goes extinct. The defining feature of this condition is its reducibility to a single scalar: the dynamical parameters enter only through their product, so the persistence question collapses onto one number. More refined spectral conditions replace  $\langle k \rangle$ with the leading eigenvalue of the adjacency matrix~\cite{wang2003}, but the reduction to a scalar is preserved.

Many real spreading processes, however, do not follow simple contagion dynamics. Behavior change, opinion adoption, technology uptake, and coordinated collective action typically require reinforcement from multiple independent sources before an individual adopts~\cite{centola2007}. This phenomenon, called \emph{complex contagion}, was formalized by Granovetter~\cite{granovetter1978} in the sociological setting and studied from a network theoretic perspective by Watts~\cite{watts2002}, who introduced the fractional threshold model: a node activates when the fraction of its activated neighbors exceeds a threshold $\theta$. Watts derived an analytical cascade condition for deterministic, monotone dynamics on random networks, showing that global cascades require a window of vulnerability in the joint space of mean degree and threshold.

Subsequent work generalized the framework in several directions. Dodds and Watts~\cite{dodds2004,dodds2005} introduced stochastic per-contact transmission, dose accumulation, and memory into the contagion model, identifying three universality classes of phase transitions depending on the dose-response function. Gleeson~\cite{gleeson2007,gleeson2008} developed analytical cascade-size predictions on tree-like networks. Centola~\cite{centola2010} provided experimental evidence that complex contagion spreads more effectively on clustered networks than on random ones, confirming the theoretical predictions. Miller~\cite{miller2016} analyzed hybrid phase transitions in the stochastic-transmission Watts threshold model on configuration-model networks, deriving conditions under which cascades exhibit discontinuous jumps.

A common feature of these studies is their focus on the monotone regime (once a node activates, it remains active permanently), i.e. on SIR-like dynamics where recovered nodes are removed. The SIS variant, in which recovered nodes return to the susceptible state and can be reinfected, introduces qualitatively different dynamics: reinfection cycling, the possibility of sustained endemic equilibria, and the question of whether the epidemic persists indefinitely or goes extinct in the long run. This regime was studied analytically by Dodds and Watts~\cite{dodds2005} in a mean-field formulation, which identified three classes of long-time behavior under recovery to susceptibility. On explicit networks, however, the extinction-persistence boundary of the Watts-threshold SIS model in the joint parameter space $(\beta,\theta,d)$ remains unmapped despite the relevance of this setting to recurrent social contagion.

We address this gap by numerically reconstructing the extinction-persistence boundary on both Erd\H{o}s--R\'enyi (ER)~\cite{erdos1959} and Barab\'asi--Albert (BA)~\cite{barabasi1999} networks. Using an adaptive Delaunay-based sampling strategy to concentrate computational effort near the critical surface, followed by weighted logistic regression on over 180,000 Monte Carlo trials, we reconstruct the phase boundary and characterize its geometry, sharpness, parameter dependence, and sensitivity to network topology. The resulting phase diagram provides a quantitative map of epidemic feasibility under complex contagion with recovery.

% ================================================================
\section{Model}
\begin{figure}[t]
  \centering
  \includegraphics[width=0.5\textwidth]{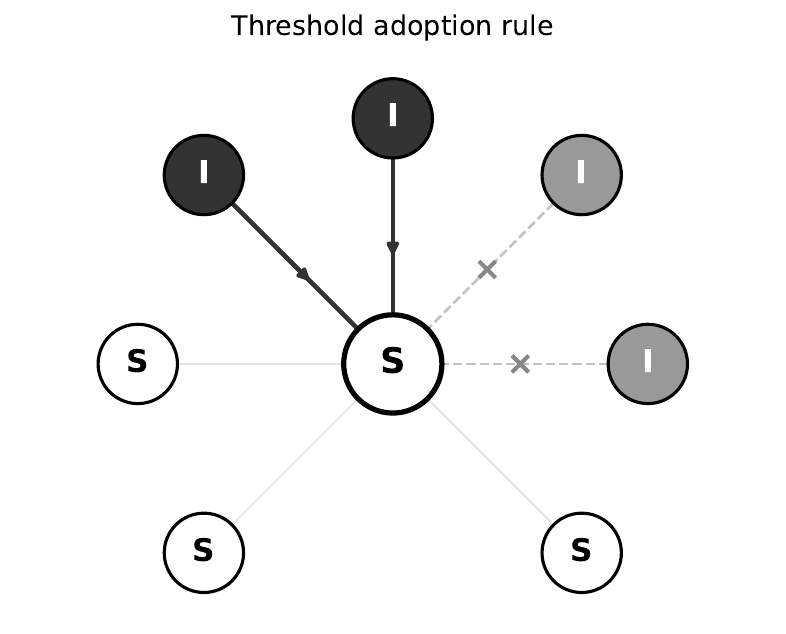}
  \caption{Illustration of the threshold adoption rule. A focal
  susceptible node with degree $k = 7$ and threshold $\theta = 0.4$
  requires at least $\lceil 0.4 \times 7 \rceil = 3$ simultaneous
  successful transmissions from infected neighbors to become infected.
  In this example, only 2 of 4 infected neighbors transmit successfully; the threshold is not met and the node remains susceptible.}
  \label{fig:model}
\end{figure}

We simulate a discrete-time stochastic SIS process on a static network augmented with a fractional adoption threshold. At each time step, every infected node independently attempts to transmit to each of its susceptible neighbors, with each attempt succeeding independently with probability~$\beta$. A susceptible node with degree~$k$ transitions to the infected state only if the number of neighbors that successfully transmit to it in that time step reaches at least $\lceil \theta k \rceil$; otherwise the transmissions have no effect and the node remains susceptible (see figure~\ref{fig:model}). When $\theta = 0$ the model reduces to a standard SIS process (any single successful transmission suffices), and increasing~$\theta$ progressively raises the barrier for adoption.

Each infected node recovers to the susceptible state after a fixed infectious duration~$d$, measured in simulation time steps. A fraction $p_0 = 0.01$ of nodes are randomly seeded as infected at $t = 0$. Because adoption requires $\lceil \theta k \rceil$ simultaneous transmissions, a single infected node generically cannot trigger activation when $\theta > 0$; persistence therefore depends on the initial infected fraction, and the boundaries reported below are conditional on the seed density $p_0 = 0.01$. Each simulation run continues until either the epidemic goes extinct (no infected nodes remain) or a maximum time horizon~$T_{\max}$ is reached. The outcome of each run is binary: extinction or persistence.

Two network topologies are studied. The first is an Erd\H{o}s--R\'enyi
(ER) random graph~\cite{erdos1959}, representing a homogeneous degree
distribution. The second is a Barab\'asi--Albert (BA) preferential
attachment network~\cite{barabasi1999}, representing a heterogeneous,
heavy-tailed degree distribution. Both networks have $N = 50\,000$ nodes
and mean degree $\langle k \rangle = 100$. The average degree is chosen so that the
integer adoption threshold $\lceil \theta k \rceil$ varies finely as
$\theta$ ranges over $[0.01, 0.30]$, while remaining of the order of an
individual's active social circle. Each network is generated once per
topology, and all simulations for a given topology use the same
realization. The parameter space spans $\beta \in [0.2, 0.9]$, $\theta \in [0.01, 0.30]$, and $d \in [2.0, 6.0]$. Each parameter configuration was simulated with $R=30$ replicates, sufficient for the per-scenario extinction probability to stabilize (standard error below $0.05$).

% ================================================================
\section{Methods}

\subsection{Adaptive Delaunay Boundary Refinement}
\label{sec:delaunay}

The extinction-persistence boundary occupies a thin manifold in the three-dimensional parameter space. Uniform grid sampling would allocate most of the computational budget to the large homogeneous regions where the outcome is deterministic (always extinct or always persistent). We therefore followed an adaptive sampling strategy based on Delaunay triangulation to concentrate sampling effort near the critical surface.

Starting from a coarse uniform grid in $(\beta, \theta, d)$, each scenario is classified into one of three regimes based on its observed extinction probability $\hat{p}_{\mathrm{ext}} = n_{\mathrm{extinct}} / R$: \emph{persistence} ($\hat{p}_{\mathrm{ext}} = 0$), \emph{extinction} ($\hat{p}_{\mathrm{ext}} = 1$), or \emph{transition} ($0 < \hat{p}_{\mathrm{ext}} < 1$). The coordinates of all simulated points are normalized to $[0,1]^3$ using the coarse-grid bounds, and the Delaunay triangulation of the point cloud is computed.

We then identify all edges of the triangulation that connect an extinction-regime vertex to a persistence-regime vertex. These edges necessarily straddle the boundary: the critical surface must pass through each such edge somewhere between its endpoints. For each identified edge, a new candidate point is placed at its midpoint in the original (unnormalized) parameter coordinates, simulated with $R = 30$ eplicates, and added to the accumulated point cloud. The Delaunay triangulation is recomputed and the process iterates, progressively concentrating sampling density near the transition manifold.

\begin{figure}[t]
  \centering
  \includegraphics[width=\textwidth]{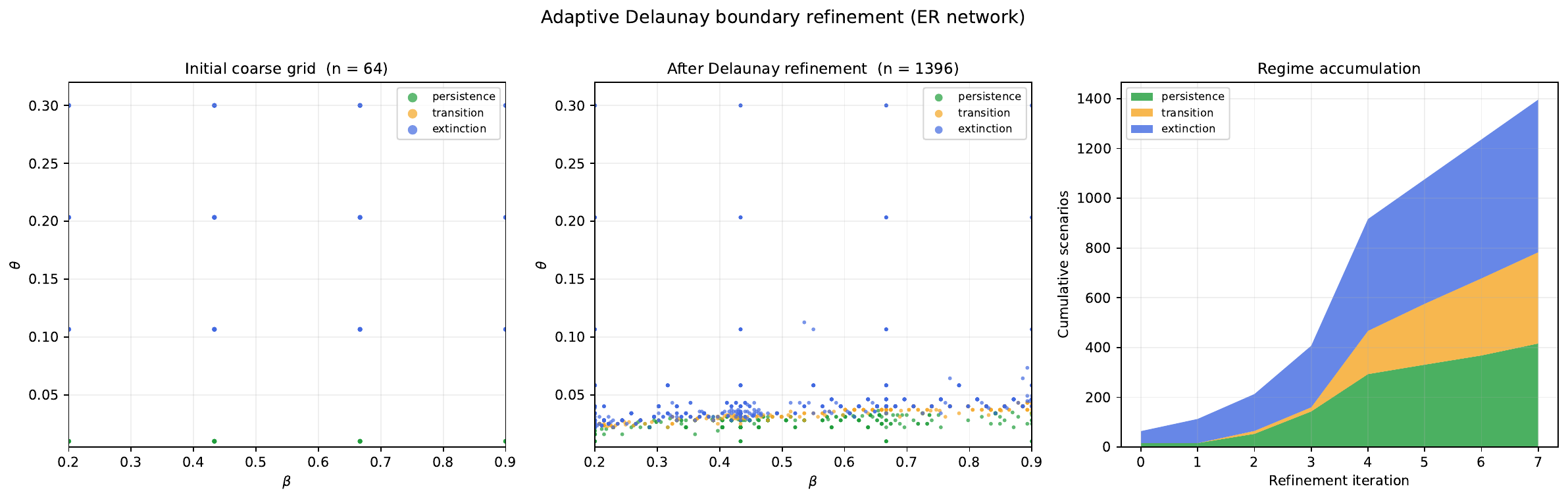}
  \caption{Adaptive Delaunay boundary refinement (ER network). Left:
  initial coarse grid. Center: point cloud after 7 iterations, showing
  enrichment near the transition region. Right: cumulative regime
  distribution across iterations.}
  \label{fig:sampling}
\end{figure}

After a few iterations, the number of boundary-straddling edges grows rapidly. To maintain a feasible computational budget, candidates are randomly subsampled from iteration~5 onward. Seven refinement iterations produce 3853 scenarios for ER and 2223 scenarios for BA. Figure~\ref{fig:sampling} illustrates the sampling enrichment near the boundary and the regime accumulation across iterations.

\subsection{Boundary Reconstruction via Weighted Logistic Regression}

The phase boundary is reconstructed by fitting a logistic regression model to the Monte Carlo outcomes. Each of the $R = 30$ replicates per scenario is treated as an independent Bernoulli trial (survived $= 1$, extinct $= 0$), and scenario-level replicate counts enter the binomial
likelihood as frequency weights. This preserves all information from every trial without discarding observations or rebalancing across regimes. Predictors $(\beta, \theta, d)$ are $z$-scored prior to fitting to ensure comparable coefficient scales.

Three nested model tiers are compared: an additive model with independent parameter effects (3~d.f.), an interaction model incorporating all pairwise products (6~d.f.), and a quadratic model
adding squared terms (9~d.f.). Model selection is performed using five-fold stratified cross-validation with the scenario-level Brier score as the evaluation metric. Stratification ensures that each fold contains a representative mix of persistence, transition, and extinction scenarios.

Table~\ref{tab:modelsel} reports the cross-validated Brier scores for both topologies. The additive model is clearly inadequate for both networks, confirming that the governing parameters do not act independently. The interaction model substantially outperforms the additive model, reducing the Brier score by more than half. The quadratic model provides negligible additional improvement on ER (0.021 vs.\ 0.022) and degrades performance on BA (0.062 vs.\ 0.040), indicating overfitting of the squared terms on the smaller BA dataset. We retain the interaction model for both topologies, providing a consistent six-parameter representation of the boundary:
\begin{equation}
  \mathrm{logit}\, P(\text{survival})
  = \beta_0 + \beta_1 \beta_z + \beta_2 \theta_z + \beta_3 d_z
  + \beta_4 \beta_z \theta_z + \beta_5 \beta_z d_z
  + \beta_6 \theta_z d_z \,.
  \label{eq:interaction}
\end{equation}
The extinction--persistence boundary is defined as the isosurface where
$P(\text{extinction}) = 0.5$.

\begin{table}[t]
  \centering
  \caption{Model selection: 5-fold stratified cross-validated Brier
  score. Lower is better. Best model per topology in bold.}
  \label{tab:modelsel}
  \begin{tabular}{lccc}
    \toprule
    Model & d.f. & ER & BA \\
    \midrule
    Additive    & 3 & 0.047 & 0.058 \\
    Interaction & 6 & \textbf{0.022} & \textbf{0.040} \\
    Quadratic   & 9 & 0.021 & 0.062 \\
    \bottomrule
  \end{tabular}
\end{table}

% ================================================================
\section{Results}

\subsection{Boundary Geometry and Transition Sharpness}

\begin{figure}[t]
  \centering
  \includegraphics[width=\textwidth]{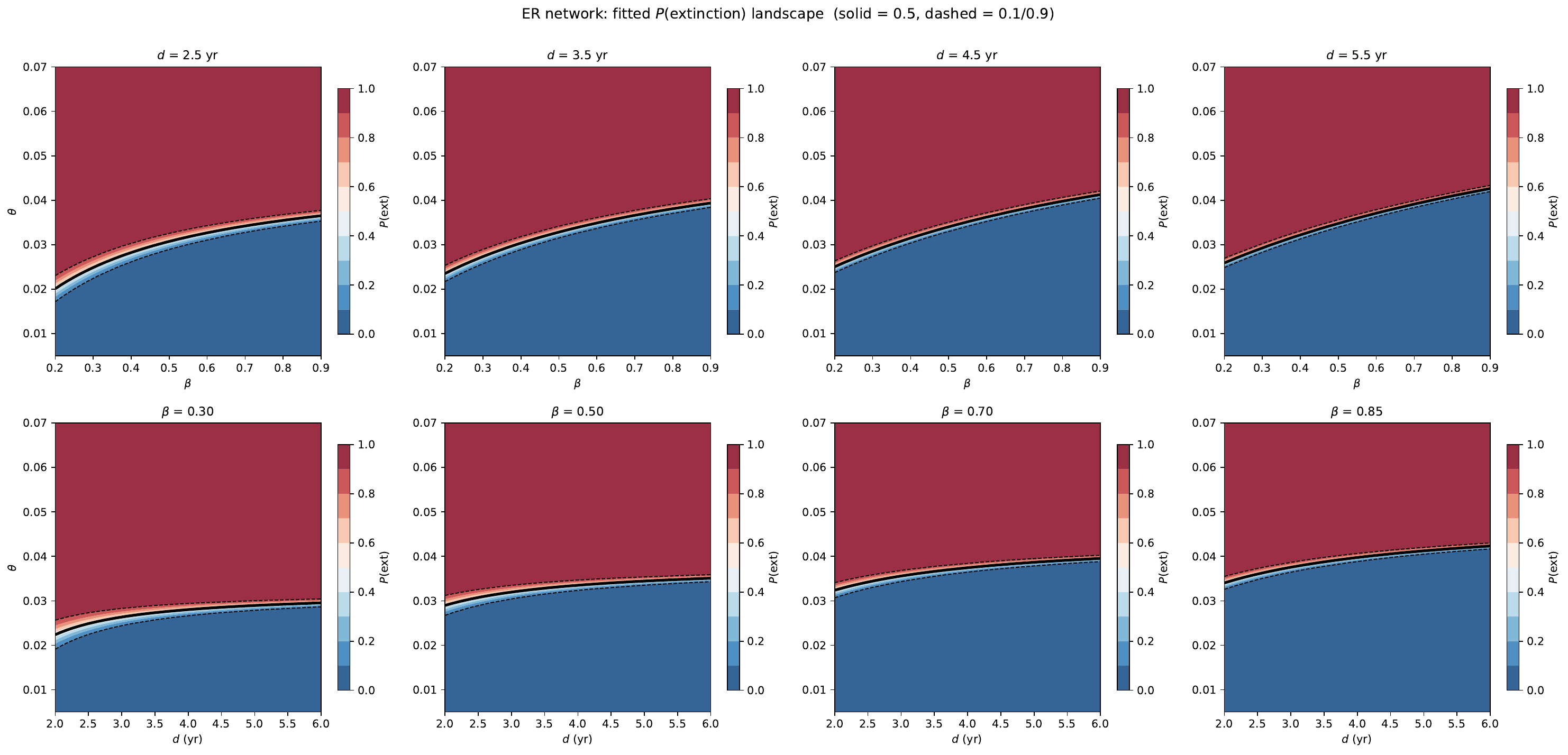}
  \caption{ER network extinction-probability landscape. Top row:
$(\beta, \theta)$ plane at fixed~$d$. Bottom row: $(d, \theta)$ plane
at fixed~$\beta$. Solid contour: $P(\mathrm{ext})=0.5$; dashed: 0.1
and 0.9.}
  \label{fig:heatmaps}
\end{figure}

Figure~\ref{fig:heatmaps} displays the extinction-probability landscape in two-dimensional cross-sections of the parameter space for the ER network. In the $(\beta, \theta)$ plane at fixed~$d$ (top row), the $P(\text{ext}) = 0.5$ boundary is a smooth, monotonically increasing curve: higher transmission rates permit higher adoption thresholds while still sustaining the epidemic. In the $(d, \theta)$ plane at fixed $\beta$ (bottom row), the boundary rises more gently with duration.

In all panels, the 10\%--90\% extinction-probability band (dashed contours) is narrow, typically spanning only $\Delta\theta \approx 0.005$--$0.008$. This indicates a sharp phase transition rather than a smooth crossover.

\subsection{Parameter Hierarchy}

Table~\ref{tab:coefficients} reports the interaction-model coefficients in raw-variable form, ranked by absolute magnitude, for both topologies. All terms are highly statistically significant ($|z| > 56$, $p \approx 0$ for all terms on both networks).

\begin{table}[t]
  \centering
  \caption{Interaction-model coefficients in raw-variable form, ranked
  by $|\mathrm{coef}|$. The response variable is $\mathrm{logit}\,P(\text{survival})$.}
  \label{tab:coefficients}
  \begin{tabular}{clrrc}
    \toprule
    Rank & Term & ER & BA & Sign \\
    \midrule
    1 & $\beta \times \theta$  & $-1558$ & $-1110$ & $-$ \\
    2 & $\theta$               & $+720$  & $+461$  & $+$ \\
    3 & $\theta \times d$      & $-463$  & $-272$  & $-$ \\
    4 & $\beta$                & $+39$   & $+36$   & $+$ \\
    5 & intercept              & $-26$   & $-20$   & $-$ \\
    6 & $\beta \times d$       & $+14$   & $+7$    & $+$ \\
    7 & $d$                    & $+11$   & $+8$    & $+$ \\
    \bottomrule
  \end{tabular}
\end{table}

Two features of this table merit attention. First, the coefficient ranking is identical across both topologies: the $\beta \times \theta$ interaction dominates, followed by the $\theta$ main effect and the $\theta \times d$ interaction. All signs are consistent. The three threshold-involving terms (ranks~1--3) are an order of magnitude larger than the remaining terms, confirming that $\theta$ is the primary axis governing epidemic feasibility and that its interactions with~$\beta$ and~$d$ shape the boundary geometry.

Second, the ER coefficients are systematically larger in magnitude than the BA coefficients, by approximately 40\% for the dominant interaction terms. This reflects a sharper transition in the ER network. This is consistent with the regime statistics, where the ER dataset contains a smaller fraction of transition-regime scenarios (22.4\%) than BA (30.3\%), indicating a narrower stochastic band around the critical surface in homogeneous networks.

Marginal effects evaluated at the center of the transition region quantify the parameter hierarchy directly (Table~\ref{tab:marginal}). Per standard deviation of variation in the sampled space, $\theta$ is 5--9$\times$ more influential than~$\beta$ and 12--14$\times$ more influential than~$d$ on both topologies. The hierarchy $\theta \gg \beta > d$ is robust.

\begin{table}[t]
  \centering
  \caption{Marginal effects at the centre of the transition region,
  evaluated at median $(\beta, \theta, d)$ of transition-regime
  scenarios. ``Per 1 SD'' standardises by the parameter's standard
  deviation in the dataset.}
  \label{tab:marginal}
  \begin{tabular}{lrrrr}
    \toprule
    & \multicolumn{2}{c}{ER} & \multicolumn{2}{c}{BA} \\
    \cmidrule(lr){2-3} \cmidrule(lr){4-5}
    Parameter & $\partial (\text{logit }P) / \partial x$ & per 1 SD
              & $\partial (\text{logit }P) / \partial x$ & per 1 SD \\
    \midrule
    $\theta$ & $+343.4$ & $+7.61$ & $+294.9$ & $+8.53$ \\
    $\beta$  & $-6.5$   & $-1.38$ & $-4.6$   & $-0.99$ \\
    $d$      & $-0.5$   & $-0.60$ & $-0.5$   & $-0.62$ \\
    \bottomrule
  \end{tabular}
\end{table}

Both $\beta$ and $d$ exert their influence primarily through interactions with $\theta$ rather than as independent main effects. The $d$ main effect is the smallest term in the model (rank~7 in Table~\ref{tab:coefficients}), with its contribution carried almost entirely by the $\theta \times d$ interaction (rank~3); $\beta$ likewise acts predominantly through the dominant $\beta \times \theta$ term (rank~1), its main effect (rank~4) being comparatively minor. Across the sampled range $d$ is the weakest of the three control parameters, consistent with the marginal effects of Table~\ref{tab:marginal}.

\subsection{Critical Threshold Curves}

\begin{figure}[t]
  \centering
  \includegraphics[width=\textwidth]{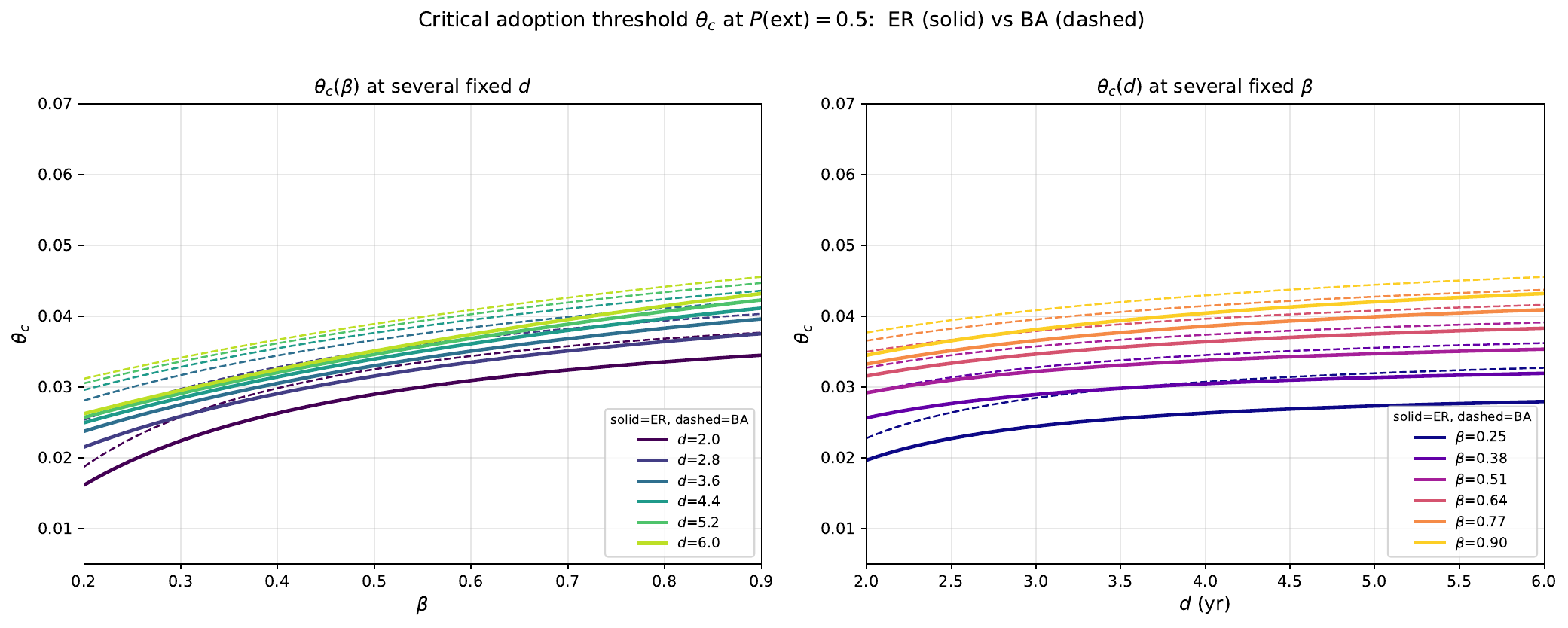}
  \caption{Critical adoption threshold $\theta_c$ at
  $P(\mathrm{ext}) = 0.5$. Left: $\theta_c$ vs.~$\beta$ at several
  fixed~$d$ (well-separated curves). Right: $\theta_c$ vs.~$d$ at
  several fixed~$\beta$ (bunched curves). Solid: ER; dashed: BA. The BA
  boundary is systematically above ER across the parameter space.}
  \label{fig:thetac}
\end{figure}

Figure~\ref{fig:thetac} makes the parameter hierarchy visible. The left panel plots the critical threshold $\theta_c$ (the maximum adoption barrier compatible with epidemic persistence) as a function of~$\beta$ for several fixed values of~$d$, for both ER~(solid) and BA~(dashed) networks. The curves are well separated: at fixed~$d$, increasing $\beta$ from 0.3 to 0.8 raises $\theta_c$ by approximately 60\%. The right panel shows $\theta_c$ as a function of~$d$ for several fixed values of~$\beta$. The curves are tightly bunched: increasing $d$ from 2 to 6~years shifts $\theta_c$ by only 15--25\% across the $\beta$ range.

\subsection{Topology Comparison}

\begin{figure}[t]
  \centering
  \includegraphics[width=\textwidth]{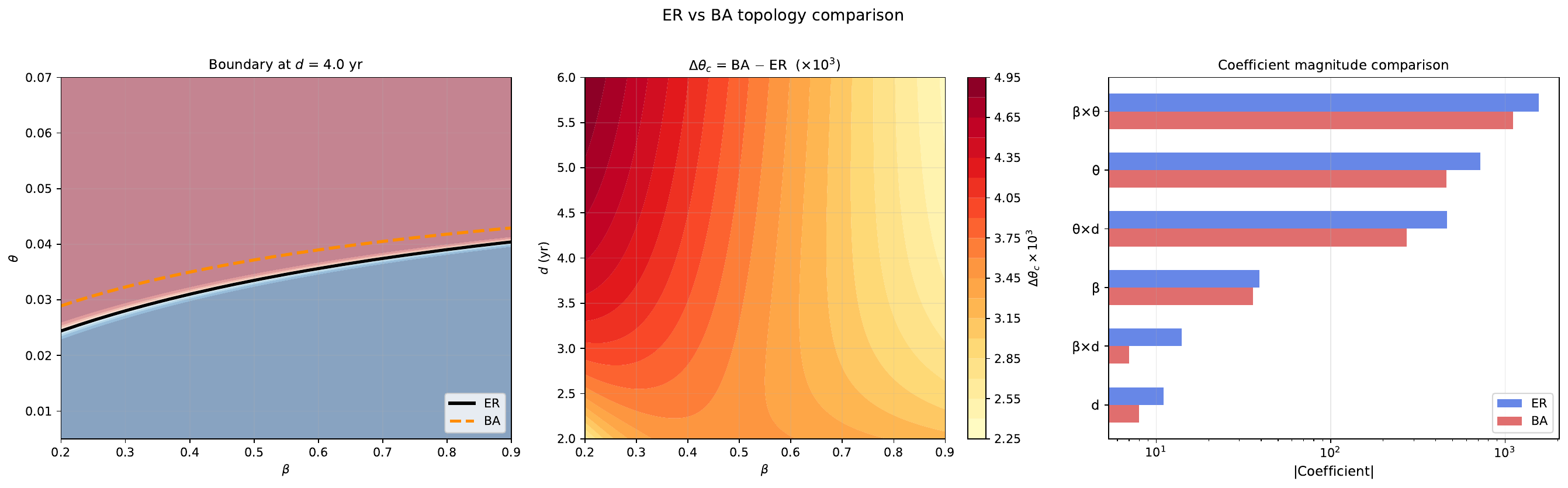}
  \caption{Topology comparison. Left: overlaid $P(\mathrm{ext}) = 0.5$
  boundaries at $d = 4.0$~yr. Centre: boundary shift
  $\Delta\theta_c = \theta_c^{\mathrm{BA}} - \theta_c^{\mathrm{ER}}$
  across $(\beta, d)$ space. Right: coefficient magnitude comparison on
  log scale.}
  \label{fig:topology}
\end{figure}

The boundary structure is qualitatively invariant across ER and BA networks: the coefficient ranking is identical, the signs are consistent, and the parameter hierarchy $\theta \gg \beta > d$ is preserved. Two quantitative differences emerge, visible in Fig.~\ref{fig:thetac} and detailed in Fig.~\ref{fig:topology}.

First, the BA network tolerates higher adoption thresholds at comparable $(\beta, d)$, with $\theta_c$ shifted upward by $\Delta\theta \approx 0.003$--$0.005$ (Fig.~\ref{fig:topology}, center panel). This is consistent with hub-facilitated activation: high-degree nodes in BA networks require fewer neighbors in absolute count to cross the threshold ($\lceil\theta k\rceil$ grows sublinearly with $k$ for small~$\theta$), and once activated, provide coordinated exposure to their many neighbors, seeding cascades unavailable in the homogeneous ER topology. This extends the known role of degree heterogeneity in facilitating simple spreading~\cite{pastor-satorras2015,newman2002,barabasi1999} to the complex-contagion setting.

Second, the ER transition is sharper, with raw coefficients approximately 40\% larger in magnitude (Fig.~\ref{fig:topology}, right panel) and a narrower stochastic band. This suggests that homogeneous networks exhibit a more abrupt switch between extinction and persistence, while the degree heterogeneity of BA networks introduces a broader transition region.

% ================================================================
\section{Discussion}
This study provides a systematic characterization of the extinction-persistence phase boundary for the stochastic Watts-threshold SIS model. The boundary plays the role that the $R_0 = 1$ condition plays in simple SIS, separating extinction from persistence, but differs from it in two respects. First, it cannot be reduced to a single scalar: the three control parameters enter as an irreducible joint condition rather than through a single product, so no one-dimensional quantity orders the diagram. Second, it is a persistence threshold from a finite seed rather than an invasion threshold: since a lone infective cannot initiate spread when $\theta > 0$, the classical criterion of one infective in a susceptible population does not apply, and the boundary location depends on the seed density.

The invariance across ER and BA topologies (same coefficient ranking, same signs, same parameter hierarchy) suggests that the dominant features of the phase diagram arise from the threshold mechanism itself rather than from topology-specific effects. The quantitative shift in boundary location (BA tolerating higher thresholds) reflects how degree heterogeneity modulates the coordination barrier but does not alter the fundamental parameter dependence.

From a practical standpoint, the parameter hierarchy $\theta \gg \beta > d$ indicates that the adoption barrier is the primary determinant of whether threshold-dependent spreading can be sustained. Among the dynamical parameters, the fitted model ranks per-contact transmission probability above infectious duration over the sampled range, suggesting that reducing transmission is the more effective intervention lever where threshold contagion governs spread; this ordering is specific to the parameter ranges studied and should not be extrapolated beyond them.

Several limitations should be acknowledged. The analysis is purely numerical; an analytical derivation of the boundary from the generating-function framework~\cite{gleeson2007,gleeson2008} or pair-approximation methods remains an important open problem. Results are conditional on single network realizations per topology, and boundary location may vary across network draws. Because persistence is measured from a fixed seed density, the boundary is conditional on that seed; under bistable dynamics its location would shift with seed size, and seed-size dependence of cascade conditions is well established for threshold models~\cite{gleeson2007,miller2016}. A systematic seed-size sweep, and tests for hysteresis, remain for future work. The threshold~$\theta$ is fixed globally; heterogeneous thresholds, as in the original Watts model~\cite{watts2002}, may modify the boundary geometry. Finally, the deterministic recovery (fixed~$d$) differs from the memoryless exponential recovery of standard Markovian SIS; stochastic recovery would alter the temporal correlation structure and may shift the boundary.

\section{Conclusion}
We have mapped the extinction-persistence boundary of the stochastic Watts-threshold SIS model across the joint parameter space $(\beta,\theta,d)$ on ER and BA networks, using adaptive sampling
concentrated near the critical surface. The boundary is governed primarily by the adoption threshold, with transmission and duration playing secondary and asymmetric roles, and its qualitative structure is invariant across the two topologies. By framing the result as a finite-seed persistence boundary rather than an invasion threshold, the characterization clarifies how complex contagion with recovery departs from the classical $R_0$ picture, and provides a quantitative baseline for analytical treatment and for extensions to empirical topologies and heterogeneous thresholds.

% ================================================================


\begin{thebibliography}{15}

\bibitem{pastor-satorras2015}
Pastor-Satorras, R., Castellano, C., Van~Mieghem, P., Vespignani, A.:
Epidemic processes in complex networks.
Rev.\ Mod.\ Phys.\ \textbf{87}, 925--979 (2015)

\bibitem{wang2003}
Wang, Y., Chakrabarti, D., Wang, C., Faloutsos, C.:
Epidemic spreading in real networks: an eigenvalue viewpoint.
In: 22nd Int.\ Symp.\ on Reliable Distributed Systems, pp.~25--34. IEEE (2003)

\bibitem{centola2007}
Centola, D., Macy, M.:
Complex contagions and the weakness of long ties.
Am.\ J.\ Sociol.\ \textbf{113}(3), 702--734 (2007)

\bibitem{granovetter1978}
Granovetter, M.:
Threshold models of collective behavior.
Am.\ J.\ Sociol.\ \textbf{83}(6), 1420--1443 (1978)

\bibitem{watts2002}
Watts, D.J.:
A simple model of global cascades on random networks.
Proc.\ Natl.\ Acad.\ Sci.\ \textbf{99}(9), 5766--5771 (2002)

\bibitem{dodds2004}
Dodds, P.S., Watts, D.J.:
Universal behavior in a generalized model of contagion.
Phys.\ Rev.\ Lett.\ \textbf{92}(21), 218701 (2004)

\bibitem{dodds2005}
Dodds, P.S., Watts, D.J.:
A generalized model of social and biological contagion.
J.\ Theor.\ Biol.\ \textbf{232}(4), 587--604 (2005)

\bibitem{gleeson2007}
Gleeson, J.P., Cahalane, D.J.:
Seed size strongly affects cascades on random networks.
Phys.\ Rev.\ E \textbf{75}(5), 056103 (2007)

\bibitem{gleeson2008}
Gleeson, J.P.:
Cascades on correlated and modular random networks.
Phys.\ Rev.\ E \textbf{77}(4), 046117 (2008)

\bibitem{centola2010}
Centola, D.:
The spread of behavior in an online social network experiment.
Science \textbf{329}(5996), 1194--1197 (2010)

\bibitem{miller2016}
Miller, J.C.:
Complex contagions and hybrid phase transitions.
J.\ Complex Netw.\ \textbf{4}(2), 201--223 (2016)

\bibitem{erdos1959}
Erd\H{o}s, P., R\'enyi, A.:
On random graphs I.
Publ.\ Math.\ Debrecen \textbf{6}, 290--297 (1959)

\bibitem{barabasi1999}
Barab\'asi, A.-L., Albert, R.:
Emergence of scaling in random networks.
Science \textbf{286}(5439), 509--512 (1999)

\bibitem{newman2002}
Newman, M.E.J.:
Spread of epidemic disease on networks.
Phys.\ Rev.\ E \textbf{66}(1), 016128 (2002)

\end{thebibliography}
\end{document}